\documentclass[a4paper,11pt]{article}
\usepackage{amsfonts}
\usepackage{latexsym}
\usepackage{amsmath,amssymb}
\usepackage{mathrsfs}
\usepackage{cite}

\def\bbc{{\Bbb C}}
\def\bbr{{\Bbb R}}
\def\bbz{{\Bbb Z}}

\newtheorem{rem}{Remark}

\newtheorem{theo}{Theorem}
\newtheorem{corol}{Corollary}

\newtheorem{ex}{Example}

\def\tr{\mathrm{tr\,}}

\def\ad{\mathrm{ad\,}}

\def\diag{\mathrm{diag\,}}
\def\rmi{\mathrm{i}}
\def\rme{\mathrm{e}}
\def\rmd{\mathrm{d\,}}

\begin{document}
\title{On a $S$-Integrable Nonlinear Evolution Equation\\ with an Additional Algebraic Constraint}
\author{Tihomir Valchev \\
\small Institute of Mathematics and Informatics,\\
\small Bulgarian Academy of Sciences,\\
\small Acad. Georgi Bonchev Str., Block 8, 1113 Sofia, Bulgaria,\\
\small tiv@math.bas.bg}
\date{}
\maketitle

\begin{abstract}
We introduce a matrix nonlinear evolution equation that obeys an additional cubic constraint. That matrix partial differential equation in two independent variables is integrable in the sense of inverse scattering transform. It has a zero curvature representation with a Lax pair being a linear bundle in pole gauge and is a generalization of Heisenberg ferromagnet equation related to the special linear algebra of arbitrary rank. We prove that the scattering operator is gauge equivalent to a linear bundle Lax operator with a complex diagonal $\lambda$-leading term. We derive a few reductions e.g. a Cartan reduction, and an orthogonal reduction of the nonlinear evolution equation under consideration.
\\[0.2cm]
\end{abstract}

\section{Introduction}\label{intro}

In \cite{bgsiam2020} the author of the present paper introduced the following $1+1$-dimensional nonlinear evolution equation 
\begin{equation}
\rmi S_t = a [S,S_{xx}] + \frac{3a}{2} \left(S^2S_{x}S\right)_x + b \left[S^2, S_{x}\right]_x 
\label{ghf_0}\end{equation}
for the function $S:\bbr^2\to\mathfrak{sl}(n,\bbc)$ that is assumed to be smooth almost everywhere. Above, "$\rmi$" is the imaginary unit, $a$ and $b$ are arbitrary complex numbers, $[.,.]$ is the 
usual matrix commutator, and subscripts denote partial differentiation. It is also required that the matrix $S(x,t)$ obeys the algebraic constraint
\begin{equation}
S^3 = S \, .
\label{constr_0}\end{equation}
It turns out that the partial differential equation (\ref{ghf_0}) is $S$-integrable. It has a Lax representation with a Lax pair of the form:
\begin{equation}
\begin{split}
L(\lambda) & =  \rmi\partial_x - \lambda S(x,t) ,\quad
A(\lambda) =  \rmi\partial_t + \lambda V_1(x,t) + \lambda^2 V_2(x,t), \\
V_2 & =  a S + b\left(S^2 - \frac{2r}{n}I\right),\qquad r = \frac{\mathrm{tr\,}(S^2)}{2}\, ,\\
V_1 & =  \mathrm{i} a [S,S_{x}] + \frac{3\mathrm{i} a}{2}S^2S_{x}S + \mathrm{i}\, b \left[S^2, S_{x}\right],
\end{split}
\label{lax_a}\end{equation}
where $\lambda\in\bbc$ is spectral parameter and $I$ is the unit matrix. It is easy to see that when $S(x,t)$ is an invertible matrix and $a=2$ equation (\ref{ghf_0}) is reduced to the Heisenberg ferromagnet equation  
\begin{equation}
\rmi S_t = \frac{1}{2} [S,S_{xx}], \qquad S^2 = I.
\label{hf_0}
\end{equation}
In the special case when $S$ takes values in $\mathfrak{su}(2)$ the equation (\ref{hf_0}) describes the evolution of a chain with spin $1/2$ while in the $\mathfrak{su}(3)$-case it describes the dynamics of a chain of spin 1, see \cite{BoPo90} for more details. The complete integrability of (\ref{hf_0}) was proved by Zakharov and Takhtadzhyan, see \cite{takht, zakhtakh, blue-bible} and it has been subject to numerous integrable analogues or generalizations e.g. magnet equation related to various simple Lie algebra \cite{forkul, golsok2,ma4}, multicomponent Landau-Lifshitz equation \cite{golsok}, the $2+1$-dimensional Ishimori equation \cite{ishim}, the $2+1$-dimensional Myrzakulov equations \cite{myr1,mmnl}. 

Another special case of (\ref{ghf_0}) is obtained when $a=0$ and it is also extensively studied in literature, see \cite{gmv,side9, varna2020}. The latter equation is related to Hermitian symmetric spaces from the series $\mathrm{A.III}$.

The present paper is dedicated to the following nonlinear evolution equation
\begin{equation}
\rmi S_t = a [S,S_{xx}] + b \left[S^2, S_{x}\right]_x + c \left(S^2S_{x}S\right)_x + d (SS_xS)_x,
\label{gghf_0}\end{equation}
where $a,b,c,d\in \bbc$ and $S(x,t)\in \mathfrak{gl}(n,\bbc)$. It is assumed that the matrix $S(x,t)$ fulfills the algebraic relation
\begin{equation}
S^3 = pS^2 + q S
\label{constr_1}\end{equation}
where complex numbers $p$ and $q$ obey certain natural conditions. Obviously, the matrix equation (\ref{gghf_0}) generalizes (\ref{ghf_0}) and is a new one. The former equation will further be referred to as a generalized Heisenberg ferromagnet equation. The paper is structured  as follows. Section \ref{prel} contains some basic algebraic results regarding the properties of the matrix $S(x,t)$ and the adjoint operator $\ad_S: \mathfrak{gl}(n,\bbc) \to\mathfrak{gl}(n,\bbc)$ to be used further in text. In this sense it plays an auxiliary role in the paper. Our main results are presented in the three sections to follow. In Section \ref{gen_hf} we prove that the for some special choice of the constants $a$, $b$, $c$ and $d$ the generalized Heisenberg ferromagnet equation with the cubic constraint (\ref{constr_1}) imposed on $S(x,t)$ admits a zero curvature representation associated with linear bundle Lax operators. Section \ref{gauge trans} is dedicated to finding a canonical representation of the Lax operators of equation (\ref{gghf_0}). In Section \ref{reduction} we obtain some simple $\bbz_2$-reductions of the generalized Heisenberg ferromagnet equation. Finally, our concluding remarks are to be presented in last section.

\section{Algebraic Preliminaries}\label{prel}

In this section we shall discuss some elementary but important properties of a $n\times n$ matrix $S$ with complex  entries that is subject to the constraint
\begin{equation}
S^3 = pS^2 + q S,\qquad p, q\in\bbc.
\label{constr_2}
\end{equation}
If $p=0$ and $q=1$ the constraint (\ref{constr_2}) turns into  (\ref{constr_0}) and the matrix $S$ is diagonalizable, see
\cite{bgsiam2020}. On the other hand, if $q=0$ and $p=1$ then $S$ is not necessarily diagonalizable since it could be a nilpotent matrix ($S^2 = 0$). A natural question is what conditions on $p$ and $q$ ensure that $S$ can be set into a diagonal form and what its spectrum is. The following theorem answers this question.

\begin{theo}
If the matrix $S\in\mathfrak{gl}(n,\bbc)$ fulfills the cubic constraint (\ref{constr_2}) with
\begin{equation}
q\neq 0, \qquad p^2 + 4q\neq 0
\label{pq_cond}
\end{equation}
then $S$ is a diagonalizable matrix. Its eigenvalues are among the complex numbers $0$ and
\begin{equation}
\mu_{\pm} = \frac{p \pm \sqrt{p^2 + 4q}}{2} \ \cdot
\label{mu_pm}
\end{equation}
\label{pot_diag}\end{theo}
\noindent {\bf Proof:} $S$ is a $n\times n$-matrix defined over $\bbc$ therefore there exists a Jordan basis in $\bbc^n$ for it. Let us assume $S$ is written in a Jordan basis, i.e. $S=\diag \left(J_{n_1}\left(\lambda_1\right), \ldots,J_{n_k}\left(\lambda_k\right)\right)$, where 
\begin{equation}
J_{n_j}\left(\lambda_j\right): = \left(\begin{array}{ccccc}
\lambda_j & 1         & 0         & \ldots & 0 \\
0         & \lambda_j & 1         & \ldots & 0 \\
0         & 0         & \lambda_j & \ldots & 0 \\
\vdots    & \vdots    & 0         & \ddots & 1 \\
0         & 0         & 0         & \ldots& \lambda_j
\end{array}\right)
\label{jblock}
\end{equation}
is a Jordan block of dimension $n_j$ ($n_1 + n_2 + \ldots + n_k = n$) corresponding to the eigenvalue $\lambda_j$ of $S$.  Obviously, the condition (\ref{constr_2}) looks the same in any basis so it splits into the string of relations
\begin{equation}
\left(J_{n_j}\left(\lambda_j\right)\right)^3 = p \left(J_{n_j}\left(\lambda_j\right)\right)^2 + q J_{n_j}\left(\lambda_j\right)
\label{jblock_constr}\end{equation}
for each Jordan block. In view of (\ref{jblock}) we can write
\begin{equation}
J_{n_j}\left(\lambda_j\right) = \lambda_j I_{n_j} + N_{n_j},
\label{jblock_repr}\end{equation}
where $I_{n_j}$ is the unit matrix of dimension $n_j$ and $N_{n_j}$ is the nilpotent part of $J_{n_j}\left(\lambda_j\right)$ the same dimension. After substituting (\ref{jblock_repr}) into (\ref{jblock_constr}), we derive
\begin{equation}
\lambda_j^3 I_{n_j} + 3\lambda_j^2N_{n_j} + 3\lambda_j N_{n_j}^2 + N_{n_j}^3
= p \left(\lambda_j^2 I_{n_j} + 2\lambda_jN_{n_j} + N_{n_j}^2\right) +
q\left(\lambda_j I_{n_j} + N_{n_j}\right)\, .
\label{jblock_constr2}
\end{equation} 
Since $I_{n_j}$ and all nonzero powers of $N_{n_j}$ are linearly independent, (\ref{jblock_constr2}) splits into
\begin{eqnarray}
\lambda_j^3 &=& p \lambda_j^2 + q\lambda_j,\label{jblock_constr2_l3}\\
3\lambda_j^2 &=& 2 p\lambda_j + q, \label{jblock_constr2_l2}\\
3\lambda_j &=& p, \label{jblock_constr2_l1}\\
1 &=& 0 . \label{jblock_constr2_l0}
\end{eqnarray} 
Due to the last relation the system (\ref{jblock_constr2_l3})--(\ref{jblock_constr2_l0}) is contradictory. This contradiction is a direct consequence of the assumption that $N_{n_j}^3 \neq 0$. Therefore the constraint (\ref{jblock_constr}) can be fulfilled only if $N_{n_j}^3 = 0$ which implies $n_j\leq 3$. Let us analyze relations (\ref{jblock_constr2_l3}) and (\ref{jblock_constr2_l2}). From (\ref{jblock_constr2_l3}) we see that the eigenvalue $\lambda_j$ is either $0$ or is given by (\ref{mu_pm}). If $\lambda_j = 0$ then (\ref{jblock_constr2_l3}) and (\ref{jblock_constr2_l2}) immediately give that $p = q = 0$ which contradicts the requirement (\ref{pq_cond}). In the case when $\lambda_j \neq 0$, (\ref{jblock_constr2_l3}) and (\ref{jblock_constr2_l2}) lead to the condition
\[q\left(p^2 + 4q\right) = 0 ,\]
which again contradicts (\ref{pq_cond}). This means that $N_{n_j} = 0$, i.e. we have that $n_j = 1$. Since $J_{n_j}\left(\lambda_j\right)$ is an arbitrary Jordan block of $S$, we deduce that the Jordan normal form of $S$ is diagonal. \quad$\blacksquare$

\begin{rem}
Let $r_{+}$ and $r_{-}$ are the multiplicities of the eigenvalues $\mu_{+}$ and $\mu_{-}$ of $S$ respectively. If either of those multiplicities, say $r_{-}$, is equal to zero then it is easy to convince ourselves that $S$ is a diagonal matrix of the form $\diag\left(\mu_{+}I_{r_{+}},0_{n-r_{+}} \right)$, where $I_{r_{+}}$ is the unit matrix of dimension $r_{+}$. This is a trivial case which is why we shall further require that 
\begin{equation}
r_{+}r_{-} \neq 0 .
\label{rpm_cond}\end{equation}
\end{rem}

The above theorem gives just a sufficient condition for $S$ to be a diagonalizable matrix that is not a necessary one however. For instance, if $S$ is a projector, i.e. it fulfills (\ref{constr_2})
with $q=0$ and $p=1$, then it is diagonalizable too. 

Let us remind the reader that for any Lie algebra $\mathfrak{g}$, the adjoint operator $\ad_X: \mathfrak{g} \to\mathfrak{g}$ of $X\in\mathfrak{g}$ is defined as
\[\ad_{X}(Y):= [X, Y], \qquad Y\in\mathfrak{g}\, ,\]
where $[. , .]$ stands for the Lie bracket in $\mathfrak{g}$. For $\mathfrak{g}\equiv \mathfrak{gl}(n,\bbc)$ and $S\in\mathfrak{gl}(n,\bbc)$ subject to (\ref{constr_2}) and (\ref{pq_cond}) the following statement holds true as a direct consequence of the previous theorem.
\begin{corol}
If $S\in\mathfrak{gl}(n,\bbc)$ satisfies conditions (\ref{constr_2}) and (\ref{pq_cond}) and the complex numbers $\mu_{\pm}$ are as given in (\ref{mu_pm}) then the adjoint operator $\ad_{S}$ is diagonalizable and its spectrum consists of the eigenvalues: $\pm (\mu_{+} - \mu_{-})$, $\pm \mu_{-}$, $\pm\mu_{+}$ and $0$.
\label{corol_ads1}
\end{corol}
\noindent {\bf Proof:}
According to the previous theorem $S$ can be diagonalized. For any diagonalizable matrix $S$ the adjoint operator corresponding to it is diagonalizable too. Indeed, let us assume $S$ is transformed into its diagonal form. Then the adjoint operator $\ad_{S}$ acts on an arbitrary $n\times n$ matrix $X$ as follows:
\begin{equation}
\ad_S(X)_{kl} = \left(S_{kk} - S_{ll}\right)X_{kl}\qquad k,l = 1,2,\ldots, n.
\label{ads_diag}\end{equation}
For $X$ being equal to Weyl generator $E_{ij}$, $i,j = 1,2,\ldots, n$ (i.e. we have that $\left(E_{ij}\right)_{kl} := \delta_{ik}\delta_{jl}$, $\delta_{ij}$ denotes the Kronecker symbol) equation (\ref{ads_diag}) gives
\[\ad_S\left(E_{ij}\right) = \left(S_{ii} - S_{jj}\right)E_{ij}.\]
That means the matrix of $\ad_S$ is diagonal and its diagonal elements (eigenvalues) are all the differences possible between the eigenvalues of $S$. \qquad $\blacksquare$

The operator $\ad_S$ is a root of its minimal characteristic polynomial. Taking into account Corollary \ref{corol_ads1}, we can write
\begin{equation}
\left[\ad_S^2 - \left(\mu_{+} - \mu_{-}\right)^2\right]\left(\ad_S^2 - \mu_{-}^2\right)\left(\ad_S^2 - \mu_{+}^2\right)\ad_S = 0.
\label{ads_0}\end{equation}
In view of (\ref{mu_pm}), the equation (\ref{ads_0}) goes into
\[\left(\ad_S^2 - p^2 - 4q\right)\left[\ad_S^4 - \left(p^2 + 2q\right)\ad_S^2 + q^2\right]\ad_S = 0.\]
After multiplying the factors above, one immediately gets
\[\left[\ad_{S}^6 - 2\left(p^2 + 3q\right)\ad_{S}^4 + \left(p^2 + 3q\right)^2\ad_{S}^2 - q^2\left(p^2 + 4q\right)\right]\ad_S = 0.\] 
Thus we have just proved the following statement.
\begin{corol}
The adjoint operator of a matrix $S\in\mathfrak{gl}(n,\bbc)$ satisfying conditions (\ref{constr_2}) and (\ref{pq_cond}) obeys the polynomial relation
\begin{equation}
\ad_{S}^{7} - 2\left(p^2 + 3q\right)\ad_{S}^5 + \left(p^2 + 3q\right)^2\ad_{S}^3 - q^2\left(p^2 + 4q\right)\ad_S = 0. 
\label{ads_chareq}\end{equation}	
\end{corol}

Generally speaking an adjoint operator $\ad_X$, $X\in\mathfrak{g}$ is not invertible --- it has infinitely many right inverse operators. However, the minimal characteristic equation (\ref{ads_chareq}) allows us to choose as right inverse of $\ad_S$, where $S\in\mathfrak{gl}(n,\bbc)$ obeys the conditions (\ref{constr_2}) and (\ref{pq_cond}), the following operator
\begin{equation}
\ad_S^{-1} = \frac{1}{q^2(p^2+4q)}\left[\ad_S^2 - \left(p^2+ 3q\right)\right]^2\ad_S
\label{ads_inv}\end{equation}
This choice will be convenient when proving that the generalized Heisenberg ferromagnet equation (\ref{gghf_0}) admits a Lax representation.

\section{Lax Representation of the Generalized Heisenberg Ferromagnet Equation}\label{gen_hf}

As we mentioned at the beginning of the paper equation (\ref{ghf_0}) has a zero curvature representation and one can apply
inverse scattering method to integrate it, see \cite{jnmp}. In this section, we intend to investigate if the same is true for the generalized Heisenberg ferromagnet equation (GHF)
\begin{equation}
\rmi S_t = a [S,S_{xx}] + b \left[S^2, S_{x}\right]_x + c \left(S^2S_{x}S\right)_x + d (SS_xS)_x \, ,
\label{gghf}\end{equation}
where $a,b,c,d\in \bbc$ and $S(x,t)\in \mathfrak{gl}(n,\bbc)$. The matrix $S(x,t)$ is required to fulfill the algebraic relation
\begin{equation}
S^3 = pS^2 + q S, \qquad q\neq 0, \qquad p^2 + 4q\neq 0 .
\label{constr_3}
\end{equation}
It turns out that GHF is $S$-integrable too, at least for some appropriate choice of the coefficients $a$, $b$, $c$ and $d$ as it is stated by the following theorem.

\begin{theo}
The equation (\ref{gghf}) with the constraint (\ref{constr_3}) imposed on $S$ has a zero curvature representation provided the coefficients $a$, $b$, $c$ and $d$ are given by:
\begin{equation}
\begin{split}
a &= \frac{\alpha (p^2 +q) - \beta pq}{q^2}\, ,\qquad b = \frac{\beta q - \alpha p}{q^2}\, ,\\
c &= \frac{2\alpha (p^2 + 3q) - 2\beta pq}{q^2(p^2 + 4q)}\, ,\qquad d = \frac{\beta p^2q - \alpha p(p^2 + 3q)}{q^2(p^2 + 4q)}
\end{split}
\end{equation}
where $\alpha$ and $\beta$ are arbitrary complex numbers. 
\label{gghf_sint}\end{theo} 

\noindent {\bf Proof:} Let us consider the Lax pair:	
\begin{equation}
\begin{split}
L(\lambda) & =  \rmi\partial_x - \lambda S(x,t),\\
A(\lambda) & =  \rmi\partial_t + \lambda V_1(x,t) + \lambda^2 V_2(x,t),
\end{split}	
\label{lax}
\end{equation}
where $\lambda\in\bbc$ is spectral parameter and $S(x,t)\in\mathfrak{gl}(n,\bbc)$ is subject to the algebraic constraint
(\ref{constr_3}). We shall prove that the compatibility condition
\begin{equation}
[L(\lambda),A(\lambda)] = 0
\label{comp_cond}
\end{equation}
of the operators (\ref{lax}) leads to equation (\ref{gghf}). 

Since the condition (\ref{comp_cond}) holds identically in the spectral parameter, it splits into the following recurrence relations: 
\begin{eqnarray}
&&[S,V_{2}] = 0, \label{l3}\\
&&\partial_x V_{2}  + \rmi [S, V_{1}] = 0, \label{l2}\\
&& \partial_tS + \partial_x V_{1} = 0. \label{l1}
\end{eqnarray}
By resolving those relations we will find $V_1$ and $V_2$ in terms of $S$ and its $x$-derivative.
From the relation (\ref{l3}) and the constraint (\ref{constr_3}) it follows that we can pick up $V_2$ to be
\begin{equation}
V_2 = \alpha S + \beta S^2 , 
\label{a2}
\end{equation}
where $\alpha$ and $\beta$ are some complex-valued functions of $x$ and $t$. After substituting the quadratic expression
(\ref{a2}) into the recurrence relation (\ref{l2}), we obtain
\begin{equation}
\alpha_{x}\, S + \alpha\, S_{x} + \beta_{x}\, S^2 + \beta\, \left(S^2\right)_{x}
+ \rmi [S, V_{1}] = 0.
\label{l2a}
\end{equation}
The structure of (\ref{l2a}) hints at representing the matrix coefficient $V_1(x,t)$ as follows:
\begin{equation}
V_1 = V_1^{\mathrm{c}} + V_1^{\mathrm{r}},\qquad\left[S, V_1^{\mathrm{c}}\right] = 0.
\label{a1_split}
\end{equation}
Such a splitting is not unique however --- it is clear that that both $V_1^{\mathrm{c}}$ and $V_1^{\mathrm{r}}$ are determined up to some $S$-commuting terms. In order to fix this ambiguity, we shall further require that $V_1^{\mathrm{r}}$ obeys the condition
\[P_{S}V_1^{\mathrm{r}} = V_1^{\mathrm{r}},\]
where $P_S$ is a projector defined through the relation
\[P_{S}:=\ad^{-1}_S\ad_{S} \, .\]
Above the right inverse operator $\ad^{-1}_S$ is the same as the one introduced in (\ref{ads_inv}).
Let us now apply $P_S$ to both hand-sides of equation (\ref{l2a}). After performing some simple but long and somewhat tedious calculations, we can convince ourselves that for any $X\in\mathfrak{gl}(n,\bbc)$ we have 
\begin{equation}
P_{S}[S,X] = [S,X],\qquad P_{S}S_x = S_x,\qquad P_{S}\left(S^{2}\right)_x = \left(S^{2}\right)_x.
\label{ps_auxrel}\end{equation}
In view of (\ref{ps_auxrel}) the recurrence relation (\ref{l2a}) splits into:
\begin{eqnarray}
\alpha_x S + \beta_x S^2 = 0,\label{l2b}\\
\alpha S_x + \beta\left(S^2\right)_x + \rmi [S, V_{1}^{\mathrm{r}}] = 0.\label{l2c}
\end{eqnarray}
Let us consider equation (\ref{l2b}) first. After evaluating the trace of both hand-sides of (\ref{l2b}), we derive
\begin{equation}
\alpha_{x} \tr S + \beta_{x} \tr\left(S^2\right) = 0.
\label{ab_hom1}\end{equation}
On the other hand, after multiplying both hand-sides of (\ref{l2b}) by $S$, evaluating the trace and making use of the cubic constraint (\ref{constr_3}),
we obtain
\begin{equation}
\alpha_{x} \tr\left(S^2\right) + \beta_{x} \left[p\, \tr\left(S^2\right) + q\, \tr S\right] = 0.
\label{ab_hom2}\end{equation}
Equations (\ref{ab_hom1}) and (\ref{ab_hom2}) form a linear homogeneous system. Taking into account the conditions (\ref{pq_cond}) and (\ref{rpm_cond}), the determinant
\[\left |\begin{array}{cc}
\tr S & \tr\left(S^2\right) \\
 \tr\left(S^2\right) & p\, \tr\left(S^2\right) + q\, \tr S
\end{array}\right | = - q\left(p^2 + 4q\right)r_{+}r_{-} \]
is not zero. Therefore (\ref{ab_hom1}) and (\ref{ab_hom2}) have a unique solution
\[\alpha_x = 0, \quad \beta_x = 0 \qquad\Rightarrow\qquad \alpha = \alpha(t),\quad \beta = \beta(t).\] 
Further on we shall restrict ourselves with case when $\alpha$ and $\beta$ are just complex numbers.

Let us now consider equation (\ref{l2c}). The latter allows us to express $V^{\mathrm{r}}_1$ through $S$ and $S_x$ by using (\ref{ads_inv}) to invert the commutator. Skipping the calculations that are
rather involving, we obtain the following result: 
\begin{equation}
V_1^{\mathrm{r}} = \rmi \left(a\, [S,S_x] + b\, [S^2,S_x] + c\, S^2S_xS + d\, SS_xS\right),
\label{a1r}
\end{equation}
where we introduced the auxiliary notation:
\begin{equation}
\begin{split}
a &= \frac{\alpha (p^2 +q) - \beta pq}{q^2}\, ,\qquad b = \frac{\beta q - \alpha p}{q^2}\, ,\\
c &= \frac{2\alpha (p^2 + 3q) - 2\beta pq}{q^2(p^2 + 4q)}\, ,\qquad d = \frac{\beta p^2q - \alpha p(p^2 + 3q)}{q^2(p^2 + 4q)}\, \cdot
\end{split}
\end{equation}
After substituting (\ref{a1r}) into the recurrence relation (\ref{l1}), we derive the equation
\[\rmi V_{1,x}^{\mathrm{c}} - \left(a\, [S,S_x] + b\, [S^2,S_x] + c\, S^2S_xS + d\, SS_xS\right) + \rmi S_{t} = 0\]
which splits into a $S$-commuting part 
\[(\mathrm{id} - P_{S})V_{1,x}^{\mathrm{c}} = 0,\]
and a noncommuting one
\begin{equation}
\rmi S_t + \rmi P_{S}(V_{1,x}^{\mathrm{c}}) = a\, [S,S_{xx}] + b \left[S^2, S_{x}\right]_x + c \left(S^2S_{x}S\right)_x + d\, (SS_xS)_x \, .
\label{nee}\end{equation}
Above $\mathrm{id}$ denotes the identity operator. Finally, setting $V_1^{\mathrm{c}} = 0$, we get the generalized Heisenberg ferromagnet equation (\ref{gghf}).
This proves the theorem. \qquad $\blacksquare$

In the next section we are going to consider the Lax representation of the equation (\ref{gghf}) itself and discuss one of its
essential properties --- the canonical form.

\section{Canonical Representation of the Lax Pair of GHF}\label{gauge trans}

The present section is dedicated to a canonical representation of the Lax pair
\begin{eqnarray}
L(\lambda) & = & \rmi\partial_x - \lambda S(x,t),\label{lax1}\\
A(\lambda) & = & \rmi\partial_t + \lambda V_1(x,t) + \lambda^2 V_2(x,t),\label{lax2}
\end{eqnarray}
where $S(x,t)$ fulfills the algebraic constraint (\ref{constr_3}) and
\begin{eqnarray}
V_1 &=& \rmi \left(a\, [S,S_x] + b\, [S^2,S_x] + c\, S^2S_xS + d\, SS_xS\right),\label{v1_itro}\\
V_2 &=& \alpha S + \beta S^2,\qquad \alpha,\, \beta\in\bbc\, , \label{v2_intro}\\
a &=& \frac{\alpha (p^2 +q) - \beta pq}{q^2}\, ,\qquad b = \frac{\beta q - \alpha p}{q^2}\, ,\label{ab_intro}\\
c &=& \frac{2\alpha (p^2 + 3q) - 2\beta pq}{q^2(p^2 + 4q)}\, ,\qquad d = \frac{\beta p^2q - \alpha p(p^2 + 3q)}{q^2(p^2 + 4q)} \,\cdot\label{cd_intro}
\end{eqnarray}
Such a representation of (\ref{lax1}) and (\ref{lax2}) underlies the direct scattering problem for (\ref{gghf}) and can be used to find its special solutions and thoroughly study GHF.

Let us consider two first order differential operators of the form
\begin{equation}
L(\lambda) = \rmi\partial_x + U(x,t,\lambda),\qquad A(\lambda) = \rmi\partial_t + V(x,t,\lambda),
\label{la_gen}
\end{equation}
where functions $U$ and $V$ take values in a complex Lie algebra $\mathfrak{g}$ and depend on the (spectral) parameter $\lambda$ in an unspecified way. Let us assume that 
\begin{equation}
[L(\lambda), A(\lambda)] = 0.
\label{la_compcond}\end{equation}
In view of (\ref{la_gen}) the compatibility condition (\ref{la_compcond}) leads to the following differential interrelation between $U$ and $V$:
\begin{equation}
\rmi U_t - \rmi V_x - [U,V] = 0.
\label{zcc_gen}
\end{equation}
Clearly, the condition (\ref{la_compcond}) is preserved under the transformation
\begin{equation}
\begin{split}
L(\lambda)\to \tilde{L}(\lambda) &:= \mathfrak{G}(x,t,\lambda)L(\lambda)\left(\mathfrak{G}(x,t, \lambda)\right)^{-1},\\
A(\lambda)\to \tilde{A}(\lambda) &:= \mathfrak{G}(x,t, \lambda)A(\lambda)\left(\mathfrak{G}(x,t,\lambda)\right)^{-1}.
\end{split}
\label{gauge_tr}
\end{equation}
Above function $\mathfrak{G}$ takes values in the complex Lie group corresponding to $\mathfrak{g}$. 

The operators $\tilde{L}(\lambda)$ and $\tilde{A}(\lambda)$ can be written down as
\[\tilde{L}(\lambda) = \rmi\partial_x + \tilde{U}(x,t,\lambda),\qquad
\tilde{A}(\lambda) = \rmi\partial_t + \tilde{V}(x,t,\lambda),\]
where 
\begin{eqnarray}
\tilde{U} &=& \rmi\mathfrak{G}\partial_x\mathfrak{G}^{-1} + \mathfrak{G}U\mathfrak{G}^{-1},\label{U_gauge_tr}\\
\tilde{V} &=& \rmi\mathfrak{G}\partial_t\mathfrak{G}^{-1} + \mathfrak{G}V\mathfrak{G}^{-1}.\label{V_gauge_tr}
\end{eqnarray}
The invariance of (\ref{zcc_gen}) under (\ref{gauge_tr}) means that $\tilde{U}$ and $\tilde{V}$ obey the equation
\[\rmi \tilde{U}_t - \rmi \tilde{V}_x - [\tilde{U},\tilde{V}] = 0.\]
An important particular case of a transformation (\ref{gauge_tr}) is when $\mathfrak{G}$ does not depend on spectral parameter. Then the $\lambda$-dependence of the operators $L(\lambda)$ and $\tilde{L}(\lambda)$ (resp. $A(\lambda)$ and $\tilde{A}(\lambda)$) is the same. Following a well established tradition, we shall refer to (\ref{gauge_tr}) in this case as a gauge transformation\footnote{Sometimes one speaks of a gauge transformation in a slightly more general context, namely when it preserves the $\lambda$-dependence of the corresponding Lax operators though $\mathfrak{G}$ might depend on $\lambda$.} \cite{book, blue-bible}. The form of the pair $(U,V)$ is called a gauge. Two such pairs are termed gauge equivalent if they are connected through a gauge transformation. 

Let us specify now the $\lambda$-dependence in $U(x,t,\lambda)$ and $V(x,t,\lambda)$. We shall be interested in the case when 
\[U(x,t,\lambda) = \sum^{N_1}_{k=-M_1}U_k(x,t)\lambda^k, \qquad 
V(x,t,\lambda) = \sum^{N_2}_{l=-M_2}V_l(x,t)\lambda^{l},\]
where $U_k(x,t)\in\mathfrak{g}$, $k=-M_1,\ldots, N_1$ and $V_l(x,t)\in\mathfrak{g}$, $l=-M_2,\ldots, N_2$. Then $U(x,t,\lambda)$ and $V(x,t,\lambda)$ are transformed under the gauge transformation (\ref{U_gauge_tr}) and (\ref{V_gauge_tr}) into $\tilde{U}(x,t,\lambda)$ and $\tilde{V}(x,t,\lambda)$ which are Laurent polynomials of the same degrees, namely
\[\tilde{U}(x,t,\lambda) = \sum^{N_1}_{k=-M_1}\tilde{U}_k(x,t)\lambda^k, \qquad 
\tilde{V}(x,t,\lambda) = \sum^{N_2}_{l=-M_2}\tilde{V}_l(x,t)\lambda^{l}.\]
The coefficients $\tilde{U}_k(x,t)\in\mathfrak{g}$, $k=-M_1,\ldots, N_1$ and $\tilde{V}_l(x,t)\in\mathfrak{g}$, $l=-M_2,\ldots, N_2$ are determined from the relations:
\begin{eqnarray}
\tilde{U}_0 &=& \rmi\mathfrak{G}\partial_x\mathfrak{G}^{-1} + \mathfrak{G}U_0\mathfrak{G}^{-1},\quad
\tilde{U}_k = \mathfrak{G}U_k\mathfrak{G}^{-1},\quad k=-M_1,\ldots,-1, 1, \ldots, N_1, \label{U_gauge_tr_a}\\
\tilde{V}_0 &=& \rmi\mathfrak{G}\partial_t\mathfrak{G}^{-1} + \mathfrak{G}V_0\mathfrak{G}^{-1},\quad
\tilde{V}_l = \mathfrak{G}V_l\mathfrak{G}^{-1},\quad l = -M_2,\ldots,-1, 1, \ldots, N_2.\label{V_gauge_tr_a}
\end{eqnarray}
		
Now we are ready to formulate and prove the main result in this section regarding the linear bundle under consideration, i.e. the Lax operators in the form (\ref{lax1}) and (\ref{lax2}). 
\begin{theo}
The operators (\ref{lax1}) and (\ref{lax2}) with $S(x,t)$ that is subject to (\ref{constr_3}) and coefficients of $A(\lambda)$ as given in (\ref{v1_itro}) and (\ref{v2_intro}) are gauge equivalent to
\begin{eqnarray}
\overline{L}(\lambda) &=& \rmi\partial_x + \left[\Sigma, Q(x,t)\right] - \lambda \Sigma,\qquad
\Sigma = \diag\left(\mu_{+}I_{r_{+}}, 0, \mu_{-}I_{r_{-}}\right), 
\label{L_canon}\\
\overline{A}(\lambda) &=& \rmi\partial_t + \overline{V}_0(x,t) + \lambda\overline{V}_1(x,t) + \lambda^2\left(\alpha\Sigma + \beta \Sigma^2\right),\qquad \alpha, \beta\in\bbc\, . \label{A_canon}
\end{eqnarray}
Above $I_{r_{+}}$ and $I_{r_{-}}$ are the identity matrices of dimension $r_{+}$ and $r_{-}$ respectively, and $\mu_{+}$, $\mu_{-}$ are defined in (\ref{mu_pm}). 
\end{theo}

\noindent{\bf Proof:}	
According to Theorem \ref{pot_diag}, there is a gauge transformation $\mathfrak{G}_0(x,t)$ that diagonalizes the $\lambda$-leading term $S(x,t)$ in (\ref{lax1}). In view of (\ref{U_gauge_tr}) we can write 
\begin{equation}
\mathfrak{G}_0(x,t)S(x,t)\left(\mathfrak{G}_0(x,t)\right)^{-1} = \Sigma, \qquad
\Sigma = \diag\left(\mu_{+}I_{r_{+}}, 0, \mu_{-}I_{r_{-}}\right),
\label{s_sigma}\end{equation}
where $\mu_{\pm}$ and $0$ are the eigenvalues of $S(x,t)$ with multiplicities $r_{\pm}$ and $n-r_{-} - r_{+}$ respectively.
Taking into account (\ref{gauge_tr}), (\ref{U_gauge_tr}), (\ref{V_gauge_tr}) and (\ref{s_sigma}), the Lax operators (\ref{lax1}) and (\ref{lax2}) with the coefficients (\ref{v1_itro}) and (\ref{v2_intro}) are transformed into
\begin{eqnarray}
\tilde{L}(\lambda) &=& \rmi\partial_x + \tilde{U}_0(x,t) - \lambda\Sigma \, ,\label{lax1b}\\
\tilde{A}(\lambda) &=& \rmi\partial_t + \tilde{V}_0(x,t) + \lambda\tilde{V}_1(x,t)
+ \lambda^2 \varTheta\, , \label{lax2b}
\end{eqnarray}
where 
\begin{eqnarray*}
\tilde{U}_0(x,t) &=& \rmi\mathfrak{G}_0(x,t)\partial_x\left(\mathfrak{G}_0(x,t)\right)^{-1},\qquad
\tilde{V}_0(x,t) = \rmi\mathfrak{G}_0(x,t)\partial_t\left(\mathfrak{G}_0(x,t)\right)^{-1},\\
\tilde{V}_1(x,t) &=& \mathfrak{G}_0(x,t)V_1(x,t)\left(\mathfrak{G}_0(x,t)\right)^{-1},\qquad
\varTheta = \alpha\Sigma + \beta \Sigma^2 \, . 
\end{eqnarray*}
In complete analogy with our considerations in the proof of Theorem \ref{gghf_sint} we split the $\lambda$-free term in
(\ref{lax1b}) as follows: 
\begin{eqnarray*}
\tilde{U}_0 = \tilde{U}_0^{\mathrm{c}} + \tilde{U}_0^{\mathrm{r}}\, , \qquad \left[\Sigma, \tilde{U}_0^{\mathrm{c}}\right] = 0\, .
\end{eqnarray*}
We shall further require that
\begin{equation}
P_{\Sigma}\tilde{U}_0^{\mathrm{r}} = \tilde{U}_0^{\mathrm{r}}\, , 
\label{u0r_constr}\end{equation}
where $P_{\Sigma}$ is a projector defined through the adjoint operator $\ad_{\Sigma}$
\begin{equation}
P_{\Sigma} := \ad_{\Sigma}^{-1}\ad_{\Sigma} = \frac{1}{q^2\left(p^2 + 4q\right)}\left[\ad_{\Sigma}^3 - \left(p^2+ 3q\right)\ad_{\Sigma}\right]^2 .
\label{p_sigma}
\end{equation}
In view of (\ref{u0r_constr}) there exist $Q(x,t)\in\mathfrak{gl}(n,\bbc)$ such that
\[\tilde{U}_0^{\mathrm{r}} = \left[\Sigma, Q(x,t)\right]\, . \]
If $\tilde{U}_0^{\mathrm{c}}$ is equal to zero then the theorem is proved. If not, we need to apply a second gauge transformation
\begin{equation}
\mathfrak{G}_1(x,t) = \rme^{\rmi \mathcal{F}(x,t)}
\label{g1}\end{equation}
to the operators (\ref{lax1b}) and (\ref{lax2b}). Above $\mathcal{F}(x,t)$ is a complex $n\times n$-matrix that is required to commute with $\Sigma$. This ensures that the $\lambda$-leading terms in (\ref{lax1b}) and (\ref{lax2b}) are invariant under the transformation $\mathfrak{G}_1(x,t)$.

As a result of the action of the gauge transformation (\ref{g1}) onto (\ref{lax1b}) and (\ref{lax2b}), we have
\begin{eqnarray}
\overline{L}(\lambda) &=& \rmi\partial_x + \overline{U}_0(x,t) - \lambda\Sigma ,\label{lax1c}\\
\overline{A}(\lambda) &=& \rmi\partial_t + \overline{V}_0(x,t) + \lambda\overline{V}_1(x,t)
+ \lambda^2 \varTheta , \label{lax2c}
\end{eqnarray}
where
\begin{eqnarray*}
\overline{U}_0(x,t) &=& \mathcal{F}_x(x,t) + \rme^{\rmi \mathcal{F}(x,t)}\tilde{U}_0(x,t)\,\rme^{-\rmi \mathcal{F}(x,t)},\\
\overline{V}_0(x,t) &=& \mathcal{F}_t(x,t) + \rme^{\rmi \mathcal{F}(x,t)}\tilde{V}_0(x,t)\rme^{-\rmi \mathcal{F}(x,t)},\\
\overline{V}_1(x,t) &=& \rme^{\rmi \mathcal{F}(x,t)}\tilde{V}_1(x,t)\rme^{-\rmi \mathcal{F}(x,t)}.
\end{eqnarray*}
We require that the block-diagonal part of $\overline{U}_0(x,t)$ is equal to zero, i.e. the latter satisfies the condition
\[P_{\Sigma}\overline{U}_0 = \overline{U}_0 \, . \]
That requirement means $\mathcal{F}(x,t)$ obeys the differential equation
\[\mathcal{F}_x + \tilde{U}_0^{\mathrm{c}} = 0 \, .\]
Obviously, its solution is given by 
\begin{equation}
\mathcal{F}(x,t) = - \int\tilde{U}_0^{\mathrm{c}}(x,t)\rmd x.
\label{gauge2_dex}
\end{equation}
After replacing $\overline{U}_0(x,t)$ with $[\Sigma, Q(x,t)]$ in (\ref{lax1c}), we obtain the Lax operator (\ref{L_canon}). Thus, we have just verified that there exists a gauge transform $\mathfrak{G} = \mathfrak{G}_1\mathfrak{G}_0$ that  maps (\ref{lax1}) and (\ref{lax2}) to (\ref{L_canon}) and (\ref{A_canon}), i.e. the pair (\ref{lax1}) and (\ref{lax2}) is equivalent to (\ref{L_canon}) and (\ref{A_canon}). This proves the Theorem. 
\qquad $\blacksquare$

We shall finish this section with a short remark. It is clear that (\ref{gauge2_dex}) does not determine $\mathcal{F}(x,t)$ uniquely (it is always up to a function of $t$). Therefore there are infinitely many gauge transformations mapping (\ref{lax1}) and (\ref{lax2}) to (\ref{L_canon}) and (\ref{A_canon}). This is why we may impose some additional conditions on
those gauge transformations. Let us split the coefficient $\tilde{V}_0(x,t)$ in (\ref{lax2b}) as follows:
\[\tilde{V}_0 =  \tilde{V}_0^{\mathrm{c}} + \tilde{V}_0^{\mathrm{r}},\qquad \left[\Sigma, \tilde{V}_0^{\mathrm{c}}\right] = 0, \qquad P_{\Sigma}\tilde{V}_0^{\mathrm{c}} = \tilde{V}_0^{\mathrm{r}} \, ,\]
where $P_{\Sigma}$ is the same as in (\ref{p_sigma}). In the special case when the formal integrability condition $\partial_t\tilde{U}_0^{\mathrm{c}} = \partial_x\tilde{V}_0^{\mathrm{c}}$ holds, the overdetermined system of partial differential equations
\begin{eqnarray*}
\mathcal{F}_x + \tilde{U}_0^{\mathrm{c}} &=& 0 ,\\
\mathcal{F}_t + \tilde{V}_0^{\mathrm{c}} &=& 0 
\end{eqnarray*}
admits a solution, namely 
\[\mathcal{F}(x,t) = - \int\tilde{U}_0^{\mathrm{c}}(x,t)\rmd x - \int\tilde{V}_0^{\mathrm{c}}(x,t)\rmd t \, .\]
This means there is a gauge transformation which maps (\ref{lax1}) and (\ref{lax2}) onto (\ref{L_canon}) and (\ref{A_canon})
with $\overline{V}_0(x,t) = [\Sigma, R(x,t)]$ for some complex $n\times n$-matrix $R(x,t)$.

\section{Reductions of GHF}\label{reduction}

In this section we are going to introduce certain special cases of the nonlinear equation (\ref{gghf}) which are compatible with the algebraic constraint (\ref{constr_3}) and the Lax representation (\ref{lax1})--(\ref{cd_intro}). More specifically, those special cases are deeply related to some additional algebraic conditions imposed on the Lax operators. 

Consider two partial differential operators $L(\lambda) =\rmi\partial_x + U(x,t,\lambda)$, $U(x,t,\lambda)\in\mathfrak{g}$ and $A(\lambda) = \rmi\partial_t + V(x,t,\lambda)$, $V(x,t,\lambda)\in\mathfrak{g}$ for some Lie algebra $\mathfrak{g}$ over the complex numbers. The compatibility condition $[L(\lambda), A(\lambda)] = 0$ is equivalent to the existence of a solution to the overdetermined system
\begin{eqnarray}
&&L(\lambda)\psi(x,t,\lambda) = 0, \label{sp_problem}\\
&&A(\lambda)\psi(x,t,\lambda) = 0,\label{seclinpr}
\end{eqnarray}
where $\psi:\mathbb{D}\times \mathbb{B}\to \bbc^n$, $\mathbb{D}\subset \bbr^2$ is an open domain and $\mathbb{B}\subset\bbc$.
We shall denote fundamental sets of solutions to (\ref{sp_problem}) and (\ref{seclinpr}) by capital letters. By definition any fundamental set of solutions takes values in the Lie group $\mathrm{GL}(n,\bbc)$. 

Let $\mathrm{G}_{\rm R}$ be a finite group which preserves the linear problems (\ref{sp_problem}) and (\ref{seclinpr}). In other words for any fundamental set of solutions $\Psi(x,t,\lambda)$ to (\ref{sp_problem}) and (\ref{seclinpr}) 
\[\tilde{\Psi}(x,t,\lambda):= \mathrm{K}_g\left(\Psi(\kappa^{-1}_{g}(x,t),k^{-1}_g(\lambda))\right),
\qquad g\in\mathrm{G}_{\mathrm{R}}\]
represents another fundamental set of solutions. Above $\kappa_g:\bbr^2\to\bbr^2$ stands for a smooth mapping, $k_g$
is a conformal mapping in the complex plane, and $\mathrm{K}_g$ is a group automorphism of the group $\mathrm{GL}(n,\bbc)$.
The $\mathrm{G}_{\rm R}$-invariance of (\ref{sp_problem}) and (\ref{seclinpr}) reduces the number of independent entries in $U(x,t,\lambda)$ and $V(x,t,\lambda)$. Therefore the number of dynamical fields solving the nonlinear equation whose zero curvature representation is given by the Lax pair $L(\lambda)$ and $A(\lambda)$ is reduced too. Further, we shall refer to $\mathrm{G}_{\mathrm{R}}$ as a reduction group and the $\mathrm{G}_{\mathrm{R}}$-action on the set of fundamental solutions as well as on the Lax operators as reduction \cite{book, blue-bible, mikh1, mikh2,pla}. In the special but rather important case when $\mathrm{G}_{\mathrm{R}}$ does not transform the variables $x$ and $t$, the reduction will be called local. 

\begin{rem}
We would like to stress here that not any finite group that preserves the auxiliary problem (\ref{sp_problem}) is a reduction group. The invariance of the second auxiliary problem is rather essential.
\end{rem}

Let us see how the general concepts above work by considering a few specific examples. 

\begin{ex}{\bf (Pseudo-Hermitian reduction)}

Let the group $\bbz_2$ acts on the fundamental sets of solutions in the following way:
\begin{equation}
\Psi(x,t,\lambda) \to \tilde{\Psi}(x,t,\lambda) = \mathcal{H}\left[\Psi^{\dag}(\kappa x,t,\kappa\lambda^*)\right]^{-1}\mathcal{H},\qquad \kappa^2 = 1,
\label{hermit_red}
\end{equation}
where $\mathcal{H}$ is a diagonal matrix whose diagonal elements are equal to $\pm 1$, $\dag$ stands for
Hermitian conjugation, and $*$ means complex conjugation. The linear problems (\ref{sp_problem}) and (\ref{seclinpr})
are invariant under the reduction (\ref{hermit_red}) if and only if the symmetry conditions
\begin{eqnarray}
\mathcal{H}U^{\dag}(\kappa x,t,\kappa\lambda^*)\mathcal{H} & = & \kappa U(x,t,\lambda), 
\label{hermitred1}\\
\mathcal{H}V^{\dag}(\kappa x,t,\kappa\lambda^*)\mathcal{H} & = & V(x,t,\lambda)
\label{hermitred2}
\end{eqnarray}
hold true. More specifically, when
\[U(x,t,\lambda) = \sum^{N_1}_{i=-M_1}U_i(x,t)\lambda^i, \qquad V(x,t,\lambda) =
\sum^{N_2}_{j=-M_2}V_j(x,t)\lambda^j\]
the conditions (\ref{hermitred1}) and (\ref{hermitred2}) split into:
\begin{eqnarray*}
\mathcal{H}U^{\dag}_i(\kappa x,t)\mathcal{H} & = & \kappa^{1-i} U_i(x,t), \qquad i= -M_1,\ldots, N_1 \, ,\\
\mathcal{H}V^{\dag}_j(\kappa x,t)\mathcal{H} & = & \kappa^{-j} V_j(x,t),\qquad j= -M_2,\ldots, N_2 \, .
\end{eqnarray*}

The $\bbz_2$-reduction (\ref{hermit_red}) is local for $\kappa = 1$ and nonlocal otherwise. In the local case 
the matrix coefficients $U_i(x,t)$ and $V_j(x,t)$ are infinitesimal symmetries of a pseudo-Hermitian form with a canonical
representation given by the matrix $\mathcal{H}$. This fact explains why we call (\ref{hermit_red}) a pseudo-Hermitian reduction.

In the particular case of the linear bundle Lax pair (\ref{lax1}) and (\ref{lax2}) with the algebraic constraint (\ref{constr_3}) imposed, the reduction (\ref{hermit_red}) requires that the numbers $p$ and $q$ are both real and
\[\mathcal{H}S^{\dag}(\kappa x,t)\mathcal{H} = S(x,t) ,\quad \mathcal{H}V_{j}^{\dag}(\kappa x,t)\mathcal{H}
= \kappa^{-j}V_{j}(x,t), \qquad j = 1,2 \, .\]
As a result, the generalized Heisenberg ferromagnet equation looks as follows:
\[\rmi S_t = a [S,S_{xx}] + b \left[S^2, S_{x}\right]_x + c \left(S^2S_{x}S\right)_x + d (SS_xS)_x,\]
where all the parameters $a$, $b$, $c$ and $d$ are now real numbers.

\end{ex}

As we saw the pseudo-Hermitian reduction did not change the form of GHF. Next example substantially changes (simplifies) the form of the equation (\ref{gghf}).

\begin{ex}{\bf (Cartan reduction)}
	
Let us now consider the $\bbz_2$-action  
\begin{equation}
\Psi(x,t,\lambda)\to \tilde{\Psi}(x,t,\lambda) = C\Psi(x,t,-\lambda)C,\qquad
C = \diag(-I_m,I_{n-m}).
\label{cartanred}\end{equation}
on the fundamental sets of solutions to (\ref{sp_problem}) and (\ref{seclinpr}). The diagonal matrix $C$ appears in the adjoint action of the Cartan involution to define the Hermitian symmetric spaces of the series $\mathrm{A.III}$. This is why we use the name Cartan reduction for (\ref{cartanred}). 

The adjoint action of $C$ induces a $\bbz_2$-grading in the Lie algebra $\mathfrak{gl}(n,\bbc)$, namely:
\begin{eqnarray*}
\mathfrak{gl}(n,\bbc) &=& \mathfrak{gl}^0(n,\bbc) + \mathfrak{gl}^1(n,\bbc),\\
\mathfrak{gl}^{\sigma}(n,\bbc) &=& \{X\in \mathfrak{gl}(n,\bbc);\ CXC^{-1} = (-1)^{\sigma}X\},
\qquad\sigma = 0,1 \, .
\end{eqnarray*}
In view of the form of $C$ the elements of the eigen subspaces $\mathfrak{gl}^0(n,\bbc)$ and $\mathfrak{gl}^1(n,\bbc)$ have certain block structure as given below:
\begin{eqnarray*}
X= \left(\begin{array}{cc}
\star & 0 \\ 0 & \star
\end{array}\right),\qquad \mathrm{if} \qquad X\in \mathfrak{gl}^0(n,\bbc)\, ,\\
X = \left(\begin{array}{cc}
0 &	\star \\ \star & 0
\end{array}\right),\qquad \mathrm{if}\qquad X\in\mathfrak{gl}^1(n,\bbc)\, .
\end{eqnarray*}

The invariance of the problems (\ref{sp_problem}) and (\ref{seclinpr}) under (\ref{cartanred}) requires that
$U(x,t,\lambda)$ and $ V(x,t,\lambda)$ must obey the symmetry conditions
\begin{eqnarray}
CU(x,t,-\lambda)C = U(x,t,\lambda)\, , 
\label{cartanred1}\\
 CV(x,t,-\lambda)C = V(x,t,\lambda)\, .
\label{cartanred2}
\end{eqnarray}
For Laurent polynomial bundles, i.e. when we have that 
\[U(x,t,\lambda) = \sum^{N_1}_{i=-M_1}U_i(x,t)\lambda^i, \qquad V(x,t,\lambda) =
\sum^{N_2}_{j=-M_2}V_j(x,t)\lambda^j\]
(\ref{cartanred1}) and (\ref{cartanred2}) give rise to the conditions
\begin{eqnarray*}
CU_i(x,t)C &=& (-1)^{i}U_i(x,t),\qquad i = -M_1,\ldots, N_1\, ,\\
CV_j(x,t)C &=& (-1)^{j}V_j(x,t), \qquad j = -M_2,\ldots, N_2 \, .
\end{eqnarray*}
It is easily seen that
\begin{eqnarray*}
U_i(x,t)&\in&\mathfrak{gl}^{\sigma}(n,\bbc),\qquad i\equiv \sigma (\mathrm{mod}\, 2),\\
V_j(x,t)&\in&\mathfrak{gl}^{\sigma}(n,\bbc),\qquad j\equiv \sigma (\mathrm{mod}\, 2) .
\end{eqnarray*}
In the case of the Lax pair (\ref{lax1}) and (\ref{lax2}) with the cubic constraint (\ref{constr_3}) imposed, the Cartan reduction leads to the conditions $p=0$ and 
\begin{equation}
CS(x,t)C = - S(x,t),\qquad CV_j(x,t)C = (-1)^{j}V_j(x,t), \qquad j = 1,2 \, .
\label{sv_cartan}\end{equation}
Taking into account the condition (\ref{sv_cartan}) and the equation (\ref{a2}), we deduce that $\alpha = 0$. Thus the equation (\ref{gghf}) simplifies to
\begin{equation}
\rmi S_t = b \left[S^2, S_x\right]_x .
\label{ghf_gmv}
\end{equation}

In view of (\ref{sv_cartan}) $S(x,t)$ belongs to $\mathfrak{gl}^{1}(n,\bbc)$ and we may use the representation 
\begin{equation}
S(x,t) = \left(\begin{array}{cc}
0 & \mathbf{u}^T(x,t) \\ \mathbf{v}(x,t) & 0
\end{array}\right),
\label{s_blocks}\end{equation}
where the superscript $T$ stands for matrix transposition and $\mathbf{u}(x,t)$ and $\mathbf{v}(x,t)$ are complex $(n-m)\times m$-matrices to satisfy the cubic constraints
\[\mathbf{u}\mathbf{v}^T\mathbf{u} = q\mathbf{u},\qquad \mathbf{v}\mathbf{u}^T\mathbf{v} = q\mathbf{v} \]
that follow from (\ref{constr_3}). Using $\mathbf{u}(x,t)$ and $\mathbf{v}(x,t)$, the equation (\ref{ghf_gmv}) acquires the form
\begin{eqnarray*}
\rmi\mathbf{u}_t &= b\left(\mathbf{u}_x\mathbf{v}^T\mathbf{u} - \mathbf{u}\mathbf{v}^T\mathbf{u}_x\right)_x ,\\
\rmi\mathbf{v}_t &= -b\left(\mathbf{v}_x\mathbf{u}^T\mathbf{v} - \mathbf{v}\mathbf{u}^T\mathbf{v}_x\right)_x  .
\end{eqnarray*}

If we combine the Cartan reduction with the pseudo-Hermitian reduction (\ref{hermit_red}), we can find the following interrelation 
\[\mathbf{v}(x,t) = \mathcal{H}_n\mathbf{u}^*(\kappa x,t)\mathcal{H}_m \, ,\]
where $\mathcal{H}_{m}$ and $\mathcal{H}_{n}$ are diagonal matrices of dimension $m$ and $n$ respectively such that the representation $\mathcal{H} = \diag(\mathcal{H}_{m}, \mathcal{H}_{n})$ holds. Then the function $\mathbf{u}$ is a solution to the equation
\[\rmi\mathbf{u}_t = b\left[\mathbf{u}_x(x,t)\mathcal{H}_m\mathbf{u}^{\dag}(\kappa x,t)\mathcal{H}_n\mathbf{u}(x,t)
- \mathbf{u}(x,t)\mathcal{H}_m\mathbf{u}^{\dag}(\kappa x,t)\mathcal{H}_n\mathbf{u}_x(x,t)\right]_x \]
along with the cubic relation
\[\mathbf{u}(x,t)\mathcal{H}_{m}\mathbf{u}^{\dag}(\kappa x,t)\mathcal{H}_n\mathbf{u}(x,t)
= q \mathbf{u}(x,t).\]

\end{ex}

We have considered so far Lax pairs with coefficients in the general linear algebra. We shall demonstrate how Lax operators with coefficients that belong to orthogonal or symplectic algebras can be derived from those {\it generic} ones by imposing an appropriate $\bbz_2$-reduction.

\begin{ex}{\bf (Orthogonal/symplectic reduction)}
	
Let us define the $\bbz_2$-reduction:
\begin{equation}
\Psi(x,t,\lambda)\to \tilde{\Psi}(x,t,\lambda) = G^{-1}\left[\Psi^{T}(x,t,\lambda)\right]^{-1}G,
\label{orthsympl}
\end{equation}
where $G$ is a constant $n\times n$-matrix with complex entries that can be viewed a matrix of a nondegenerate
bilinear form. If it is symmetric, i.e. we have $G^T = G$, then $G$ can naturally be interpreted as a metric
while in the skew-symmetric case ($G^T = - G$) it plays the role of a symplectic form.\footnote{It is well-known that the symplectic case can be realized in even-dimensional vector spaces only, i.e. $n$ must be an even number.}

From the requirement that the linear problems (\ref{sp_problem}) and (\ref{seclinpr}) remain intact under (\ref{orthsympl}) we obtain the symmetry conditions:
\begin{eqnarray}
U^{T}(x,t,\lambda)G + GU(x,t,\lambda) = 0\, , \label{orthsumpl1}\\
V^{T}(x,t,\lambda)G + GV(x,t,\lambda) = 0\, .
\label{orthsumpl2}
\end{eqnarray}
In the case of rational bundles of the form
\[U(x,t,\lambda) = \sum^{N_1}_{i= -M_1}U_i(x,t)\lambda^i,\qquad V(x,t,\lambda) =
\sum^{N_2}_{j=-M_2}V_j(x,t)\lambda^{j}\]
the conditions (\ref{orthsumpl1}) and (\ref{orthsumpl2}) split into
\begin{eqnarray}
U^{T}_i(x,t)G + G U_i(x,t) &=& 0, \qquad i=-M_1,\ldots, N_1\, ,\label{orthsympl1a}\\
V^{T}_j(x,t)G + GV_j(x,t) &=& 0,\qquad j=-M_2,\ldots, N_2\, .\label{orthsympl2a}
\end{eqnarray}  	
It is evident that the equations (\ref{orthsympl1a}) and (\ref{orthsympl2a}) have the form of relations defining the  
infinitesimal symmetries of the afore-mentioned bilinear form. In the symmetric case $U_i(x,t)$ and $V_j(x,t)$ are elements of the orthogonal algebra $\mathfrak{o}(n,\bbc)\subset\mathfrak{gl}(n,\bbc)$ while in the skew-symmetric case $U_i(x,t)$ and $V_j(x,t)$ belong to the symplectic algebra $\mathfrak{sp}(n,\bbc)\subset\mathfrak{sl}(n,\bbc)$.

For the linear bundle in pole gauge (\ref{lax1}) and (\ref{lax2}) with the additional constraint (\ref{constr_3}) the symmetry conditions (\ref{orthsympl1a}) and (\ref{orthsympl2a}) turn into
\begin{equation}
S^{T}(x,t)G + G S(x,t) = 0, \qquad V^{T}_j(x,t)G + GV_j(x,t) = 0,\qquad j=1,2 \, ,
\label{sv_orthsympl2a}
\end{equation} 
that is $S(x,t)$ is either in $\mathfrak{o}(n,\bbc)$ (symmetric case) or in $\mathfrak{sp}(n,\bbc)$ (skew-symmetric case).
It is easy to check that $(S(x,t))^2$ (or any even degree of $S(x,t)$) belongs neither to $\mathfrak{o}(n,\bbc)$ nor to
$\mathfrak{sp}(n,\bbc)$. Indeed, taking into account the former relation in (\ref{sv_orthsympl2a}), we have
\[\left(S^2\right)^T = \left(S^T\right)^2 = \left(-GSG^{-1}\right)^2 = GS^2G^{-1}\qquad\Rightarrow\qquad
\left(S^2\right)^TG - GS^2 = 0 \, .\]
From this we deduce that $(S(x,t))^2$ is neither in $\mathfrak{o}(n,\bbc)$ (symmetric case) or in $\mathfrak{sp}(n,\bbc)$ (skew-symmetric case). Therefore $p$ must vanish in (\ref{constr_3}) and $\beta$ --- in (\ref{a2}). Thus the generalized
Heisenberg ferromagnet equation reduces to 
\[\rmi S_t = a [S,S_{xx}] + \frac{3a}{2q} \left(S^2S_{x}S\right)_x, \qquad a = \frac{\alpha}{q} \, \cdot\]

\end{ex}

\section{Conclusion}

We have introduced the nonlinear evolution equation (\ref{gghf}) related to the general linear algebra over the complex numbers that is subject to the cubic constraint (\ref{constr_3}). That new equation generalizes the classical Heisenberg ferromagnet equation (\ref{hf_0}) in a very natural way. We have showed that for certain values of the parameters $a$, $b$, $c$ and $d$ the generalized  Heisenberg ferromagnet equation admits a zero curvature representation with linear bundle Lax operators in pole gauge, see (\ref{lax1}) and (\ref{lax2}). Those Lax operators are gauge equivalent to another linear bundle of the canonical form (\ref{lax1c}) and (\ref{lax2c}). We have presented a local and a nonlocal version of a pseudo-Hermitian reduction of the equation (\ref{gghf}). The pseudo-Hermitian reduction has the same form but involves real parameters. Another example considered in the paper is a Cartan reduction which has significantly simplified the original (nonreduced) equation. This simplification is a result of the $\bbz_2$-grading induced by the reduction and is deeply-related to Hermitian symmetric spaces of the $\mathrm{A.III}$ type. Finally, we have derived an orthogonal and a symplectic reduction of the generalized Heisenberg ferromagnet equation. Those reductions have also simplified (\ref{gghf}). All this is far from being an exhaustive list of reductions possible.

Important issues that have been ignored in the present paper concern constructing special solutions to (\ref{gghf}) and developing its Hamiltonian formulation. We are going to address all those issues elsewhere.
		
We have considered a generalized Heisenberg ferromagnet equation whose Lax pair is a linear bundle in pole gauge with a quadratic flow. It is meaningful to search for generalizations whose Lax pairs depends on the spectral parameter in a more complicated way. Such examples are given by linear bundles with mixed flows or rational bundles which may be invariant under the transformation $\lambda\to 1\slash\lambda$. Such nonlinear evolution equations can be viewed integrable deformation of the equation (\ref{gghf}).

\section{Acknowledgements}
The work has been financially supported by Grant KP--06--N 62/5 of Bulgarian National Science Fund.


\begin{thebibliography}{99.}
\bibitem{BoPo90}
Borovik, A.E., Popkov, V.Yu.: Completely Integrable Spin-1 Chains,
{\it Sov. Phys. JETPH} {\bf 71}, 177--85 (1990).
\bibitem{forkul}
Fordy, A., Kulish, P., Nonlinear Schr\"{o}dinger Equations and Simple Lie
Algebras, {\it Commun. Math. Phys.} {\bf 89}, 427--443 (1983).
\bibitem{side9}
Gerdjikov, V.S., Grahovski, G.G., Mikhailov, A.V., Valchev, T.I.: Polynomial Bundles and Generalised Fourier
Transforms for Integrable Equations on A.III-type Symmetric spaces, {\it SIGMA} {\bf 7} 096, 48 pages (2011).
\bibitem{gmv}
Gerdjikov, G., Mikhailov, A., Valchev, T., Reductions of Integrable Equations
on A.III-type Symmetric Spaces, {\it J. Phys. A: Math. Theor.} {\bf 43}, 434015 (2010).
\bibitem{book}
Gerdjikov, G., Vilasi, G., Yanovski, A., {\it Integrable Hamiltonian Hierarchies
--- Spectral and Geometric Methods}, Springer, Heidelberg, 2008.
\bibitem{golsok2}
Golubchik, I.Z., Sokolov V.V., Generalised Heisenberg Equations on Z-graded Lie Algebras, {\it Theor. 
Math. Phys.} {\bf 120}, 1019--1025 (1999).
\bibitem{golsok}
Golubchik, I. Z., Sokolov, V. V., Multicomponent Generalization of the Hierarchy
of the Landau-Lifshitz Equation, {\it Theor. Math. Phys.} {\bf 124}, N. 1, 909--917 (2000).
\bibitem{ishim}
Ishimori, Y., Multi-vortex Solutions of a Two-dimensional Nonlinear Wave Equation,
{\it Prog. Theor. Phys.} {\bf 72}, 33--37 (1984).
\bibitem{ma4}
Ma, W. X., Shen, S. F., Yu, S. M., Zhang H. Q.: An Integrable so(3, R)-Counterpart of the Heisenberg Soliton Hierarchy,
{\it Rep. Math. Phys.} {\bf 74}, Issue 3, 283--299 (2014).
\bibitem{mikh1}
Mikhailov, A., Reduction in the Integrable Systems. Reduction Groups, {\it Lett. JETF} {\bf 32},
187--192 (1979).
\bibitem{mikh2}
Mikhailov, A., The Reduction Problem and Inverse Scattering Method, {\it Physica D} {\bf 3},  73--117 (1981).
\bibitem{mmnl}
Myrzakulov, R., Mamyrbekova, G., Nugmanova, G., Lakshmanan, M., Integrable (2+1) Dimensional 
Spin Model: Geometric and Gauge Equivalent Counterparts, Solitons and Coherent Structures,
{\it Phys. Lett. A} {\bf 233},  391--396 (1997).
\bibitem{myr1}
Myrzakulov, R., Vijayalarshmi, S., Nugmanova G. and Lakshmanan, M.: A (2+1) Dimensional Integrable
Spin models with Self-consistent Potentials, Symmetry {\bf 7}, 1352--1373 (2015).
\bibitem{takht}
Takhtajan, L. A., Integration of the Continuous Heisenberg Spin Chain through the Inverse Scattering Method,
{\it Phys. Lett. A} {\bf 64}, 235--237 (1977).
\bibitem{blue-bible}
Takhtadjan, L., Faddeev, L., {\it The Hamiltonian Approach to Soliton Theory}, Springer Verlag, Berlin (1987).
\bibitem{pla}
Valchev, T., On Mikhailov's Reduction Group, {\it Phys. Lett. A} {\bf 379}, 1877–1880 (2015).  
\bibitem{varna2020}
Valchev, T., Multicomponent Nonlinear Evolution Equations of the Heisenberg Ferromagnet Type.
Local Versus Nonlocal Reductions, In: {\it Geometry, Integrability and Quantization}, {\bf 22},
Eds.: I. Mladenov, V. Pulov and A. Yoshioka, Avangard Prima, Sofia, 2021, 274--285 (Proc. XXII-nd
International Conference "Geometry, Integrability and Quantization", June 8--13, 2020, Varna, Bulgaria),
arXiv: 2011.13437v1[nlin.SI].
\bibitem{bgsiam2020}
Valchev, T., A Generic Nonlinear Evolution Equation of Magnetic Type I. Reductions, In: {\it Advanced Computing in
Industrial Mathematics}, Studies in Computational Intelligence, {\bf 1076}, Eds: I. Georgiev, H. Kostadinov, E. Lilkova,
Springer, 2023, 166--177  (Proc. of 15th Annual Meeting of the Bulgarian Section of SIAM, December 15--17, 2020, Sofia,
Bulgaria).
\bibitem{jnmp}
Valchev, T., A Generic Nonlinear Evolution Equation of Magnetic Type II. Particular Solutions, {\it Jour. Nonl. Math. Phys.} {\bf 33}, 49 (2026).
\bibitem{zakhtakh}
Zakharov, V., Takhtadzhyan, L., Equivalence of the Nonlinear Schr\"odinger Equation and the
Equation of a Heisenberg Ferromagnet, {\it Theor. Math. Phys.} {\bf 38} 17-–23 (1979).

\end{thebibliography}
\end{document}